\documentclass[aps,prb,twocolumn,amsmath,superscriptaddress]{revtex4}
\usepackage{graphicx}
\usepackage{color}

\begin{document}

\title{Spin dynamics of a Mn atom in a semiconductor quantum dot under resonant optical excitation}

\author{S. Jamet}
\affiliation{CEA-CNRS group "Nanophysique et
semiconducteurs", Institut N\'eel, CNRS \& Universit\'e
Joseph Fourier, 25 avenue des Martyrs, 38042 Grenoble,
France}

\author{H. Boukari}
\affiliation{CEA-CNRS group "Nanophysique et
semiconducteurs", Institut N\'eel, CNRS \& Universit\'e
Joseph Fourier, 25 avenue des Martyrs, 38042 Grenoble,
France}

\author{L. Besombes}
\email{lucien.besombes@grenoble.cnrs.fr}
\affiliation{CEA-CNRS group "Nanophysique et
semiconducteurs", Institut N\'eel, CNRS \& Universit\'e
Joseph Fourier, 25 avenue des Martyrs, 38042 Grenoble,
France}

\date{\today}

\begin{abstract}

We analyze the spin dynamics of an individual magnetic atom
(Mn) inserted in a II-VI semiconductor quantum dot under resonant
optical excitation. In addition to
standard optical pumping expected for a resonant
excitation, we show that for particular conditions of laser
detuning and excitation intensity, the spin population can
be trapped in the state which is resonantly excited. This
effect is modeled considering the coherent spin dynamics of the coupled electronic and nuclear spin of the Mn atom optically dressed by a resonant laser field. This \emph{spin population trapping} mechanism is controlled by the combined effect of the coupling with the laser field and the coherent interaction between the different Mn spin states induced by an anisotropy of the strain in the plane of the quantum dot.

\end{abstract}

\maketitle

\section{Introduction.}

Optically controlled semiconductor quantum dots (QDs) are in many ways
similar to atomic systems. In the recent years many quantum
optics effects first demonstrated in atoms have
been observed on individual QDs. For instance, resonant optical pumping \cite{Brossel1952} has been successfully used to prepare the spin of an individual electron \cite{Atature2006}, hole \cite{Gerardot2008} or magnetic atom \cite{LeGall2010} localized in a QD. The possibility to use a strong resonant continuous wave laser field to
create hybrid matter-field states \cite{Mollow1972} and manipulate QDs
states in their solid environment has also been demonstrated in
different QD systems \cite{Jundt2008}. The Autler-Townes effect in the fine structure of a neutral or charged QD \cite{Xu2007,Xu2008}, the Mollow absorption
spectrum of an individual QD \cite{Kroner2008} and the emission of an
optically dressed exciton and biexciton complex \cite{Muller2008} have been reported. In magnetic QDs containing an individual Mn atom \cite{Besombes2004,Kudelski2007}, it has been recently
demonstrated that the energy of any spin state of a
 Mn can be tuned using the optical Stark effect induced by a
strong laser field \cite{LeGall2011}. We present here a new
way to exploit this coupling with a resonant laser field to
initialize the spin of a magnetic atom inserted in a self-assembled QD.

When Mn atoms are included in a II-VI semiconductor
self-assembled QD (CdTe in ZnTe) \cite{Wojnar2011}, the
spin of the optically created electron-hole pair (exciton)
interacts with the 5{\it d} electrons of the Mn (total spin
S=5/2). In the case of a singly Mn-doped QD, this leads to
a splitting of the once simple photoluminescence (PL)
spectrum of an individual QD into six (2S+1) components
\cite{Besombes2004,Trojnar2012,Fernandez2006}. Since the confined carriers and Mn
spin wave functions become strongly mixed, the optical
excitation of the QD affects the spin state of the Mn
through the exchange interaction offering a possibility of
optical control \cite{LeGall2009,Goryca2009,
LeGall2010,LeGall2011,Reiter2009,Reiter2012}.

The dynamics of the Mn spin at zero magnetic field is mainly
controlled by a magnetic anisotropy ${\cal D}_0$ produced by the
presence of large bi-axial strain at the Mn location \cite{LeGall2009}. This crystal field splits the spin states of
the Mn according to ${\cal D}_0S_z^2$. We demonstrated recently
the possibility to tune this fine structure using the
optical Stark effect \cite{LeGall2011}. Under strong
resonant excitation of a Mn-doped QD optical transition, optically dressed
states of the Mn atom are created. The laser induced energy shift
of the optically addressed spin state can be much larger
than the fine structure splitting of the Mn atom. The laser field can then be used to shift the
energy of any spin state of the Mn and control their
degeneracy. Here, we present experimental results showing that for particular
conditions of laser detuning and intensity, the creation of
optically dressed states in a Mn-doped QD can lead to a new
way to initialize the Mn spin that we call \emph{spin population
trapping}. We will also model and discuss the parameters that
optimize this spin preparation.

This paper is organized as follow: In section 2 we describe
a new technique for the readout of the Mn spin state based
on a two-photon resonant excitation of an individual Mn
doped QD. In section 3 we show how the resonant laser
excitation on an exciton-Mn (X-Mn) level significantly modifies the Mn spin
dynamics. In addition to
standard optical pumping expected for a resonant
excitation, we show that the spin population can
be trapped in the state which is resonantly excited. In section 4 we present a model for the spin
dynamics of an optically dressed Mn atom and show that it qualitatively explains the experimental results. We demonstrate in particular that the
Mn dynamics in the resonant optical excitation regime is strongly affected by the strain anisotropy in the QD plane.

\section{Two-photon readout of the spin state of an individual Mn atom}

The magnetic QDs used in this study are grown on a ZnTe substrate. A 6.5 monolayer thick CdTe layer is deposited at 280$^{\circ}$C by atomic layer epitaxy on a
ZnTe barrier grown by molecular beam epitaxy at
360$^{\circ}$C. The CdTe dots are formed by a high temperature
Tellurium induced process \cite{Wojnar2011} and protected
by a 300~nm thick ZnTe top barrier. The QDs are in the 10nm
wide range and a few nm high. Mn atoms are introduced
during the CdTe deposition with a density roughly equal to
the density of QDs. Non-magnetic QDs and QDs containing a
low number (1, 2,...) of Mn atoms  are then
formed.

Optical addressing of individual QDs containing magnetic
atoms is achieved using micro-spectroscopy techniques. A
high refractive index hemispherical solid immersion lens is
mounted on the surface of the sample to enhance the spatial
resolution and the collection efficiency of single dot
emission in a low-temperature ($T$=5K) scanning optical
microscope \cite{Besombes2008}.

\begin{figure}[hbt]
\includegraphics[width=2.5in]{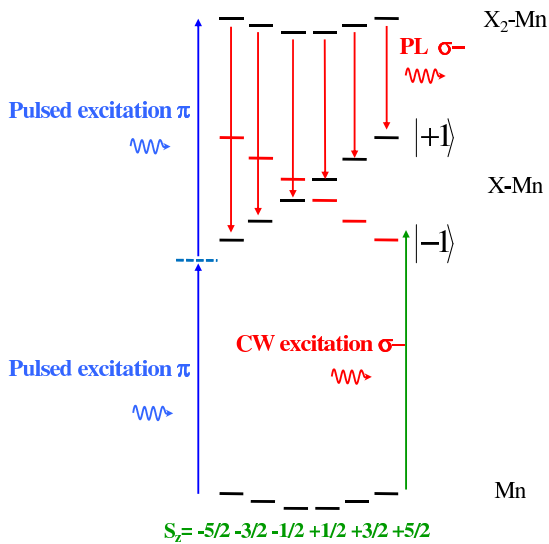}
\caption{(Color online) Scheme of the optical transitions and their
polarizations in a quantum dot containing an individual magnetic atom
and 0 (Mn), 1 (X-Mn) or 2 (X$_2$-Mn) excitons. The exciton states are split by the
exchange interaction with the Mn spin whereas in the ground (Mn)
and biexciton (X$_2$-Mn) states the energy levels results from the
fine and hyperfine structure of the Mn spin. A direct resonant excitation of the biexciton
is performed by a pulsed two-photon absorption through an intermediate virtual state whereas X-Mn states a resonantly excited by a tunable CW laser.} \label{Fig1}
\end{figure}

In order to observe the population repartition on the spin states of the Mn under resonant
optical excitation on a given X-Mn level, we developed a
technique allowing probing simultaneously the six spin states in the resonant optical excitation
regime. The principle of this experiment is presented in Fig.~\ref{Fig1}. It is based on the creation of a
 biexciton by a two-photon absorption process under resonant pulsed excitation for the Mn spin readout,
 combined with a continuous wave (CW) resonant excitation on X-Mn for the Mn spin preparation.

A picosecond pulsed laser excitation tuned between the exciton and the biexciton
levels can directly create a biexciton in a QD through a
two-photon transition \cite{Flissikowski2004}. As the
biexciton is a spin singlet state (2 paired electrons and 2 paired holes), in a first approximation it does not interact
with the Mn spin \cite{Besombes2005}. The population distribution on the six Mn
spin states is then extracted from the intensity of the PL
lines of the biexciton which is controlled by X-Mn in the final state of the biexciton recombination. As the emission of the
biexciton is shifted by 10 to 14 meV below the resonant
excitation on the X-Mn levels, it can be easily
separated from the scattered photons of the CW pumping laser.
The resonant two-photon absorption scheme used here also
avoids the injection of free carriers in the vicinity of
the QD and consequently limits the spin relaxation of the
Mn by exchange coupling with these free carriers \cite{Besombes2008}. Let us
note however that the cascade recombination of the
biexciton leaves in the QD a maximum of one exciton every 13 ns (repetition rate of the pulsed excitation). These
excitons with a characteristic X-Mn spin flip time of about 50ns \cite{LeGall2010} may slightly perturb the Mn spin preparation. However, as we will see, a
signature of the resonant preparation of the Mn is clearly observed in the biexciton signal which appears as a good probe of the Mn spin population.

\begin{figure}[hbt]
\includegraphics[width=3.3in]{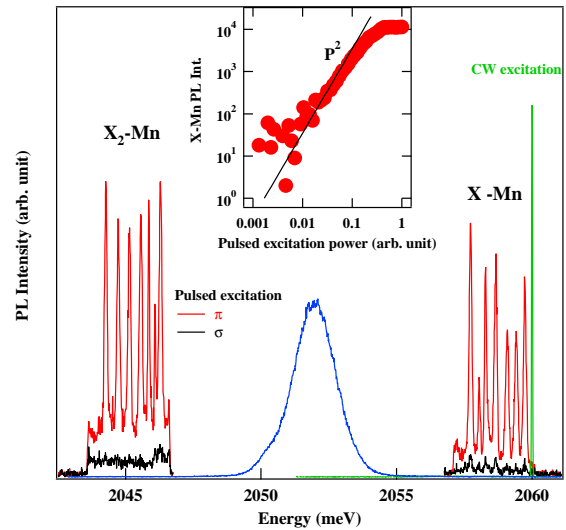}
\caption{(Color online) Photoluminescence spectra of a Mn-doped quantum
dot (QD1) obtained under a two-photon resonant
absorption of a picosecond laser pulse (blue). The resonant creation of the biexciton (X$_2$-Mn) is only
possible for a linearly polarized excitation (red)
whereas almost no photoluminescence is observed for circularly polarized pulses
(black). The influence on the Mn spin population
of a CW control laser (green) in resonance with the exciton (X-Mn) levels can be detected
in the intensity distribution of X$_2$-Mn. Inset: Photoluminescence intensity of X-Mn versus the intensity of the pulsed
linearly polarized excitation. The P$^2$ dependence is characteristic of a two-photon absorption.} \label{Fig2}
\end{figure}

The experimental evidence of the two-photon resonant formation of X$_2$-Mn is presented in Fig.~\ref{Fig2}. As
expected from optical selection rules of the two-photon transition \cite{Flissikowski2004}, PL from X$_2$-Mn can only be observed with linearly polarized laser pulses tuned in between the exciton and biexciton transitions. Excitation with circularly polarized pulses do not create
any significant QD luminescence. This confirms that one can
find experimental conditions where the PL from
the QD is dominated by the two-photon resonant excitation. The non-resonant creation of free carriers in the QD vicinity is very weak and will not perturb the Mn spin dynamics.

\begin{figure}[hbt]
\includegraphics[width=3.3in]{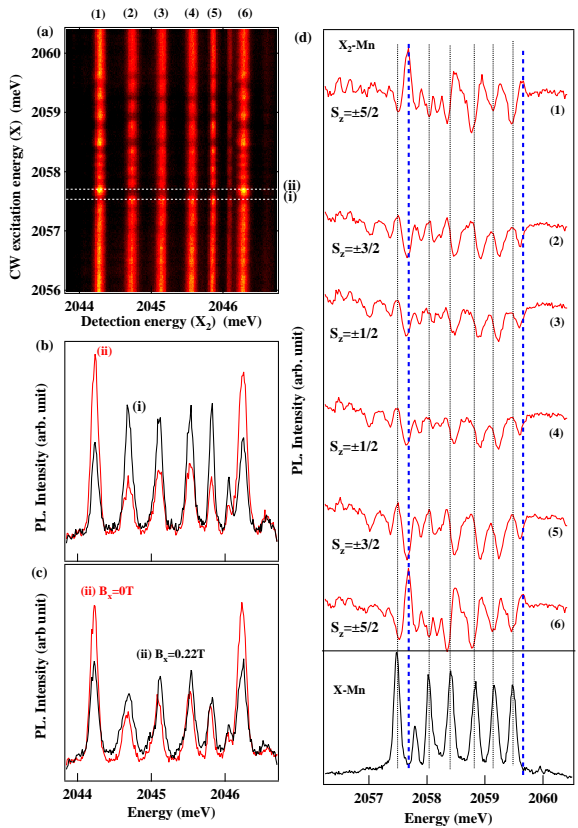}
\caption{(Color online) (a) Map of the intensity of X$_2$-Mn on QD1 versus
the energy of the linearly polarized CW resonant excitation on X-Mn for a
linearly polarized detection of the two-photon PL of X$_2$-Mn. (b) PL spectra
of X$_2$-Mn obtained for a resonant CW excitation on (i)
and (ii). (c) PL spectra of X$_2$-Mn under CW excitation on
(ii) at zero field (red) and under a transverse magnetic
field B$_x$=0.22T (black). (d) PL spectra of X-Mn (black)
and intensity of each X$_2$-Mn lines versus the CW laser
excitation energy scanned across X-Mn.} \label{Fig3}
\end{figure}

The intensity distribution of the X$_2$-Mn created by the two-photon absorption scheme presented above allows to probe
the preparation of the Mn spin by the CW control laser near the resonance with X-Mn states. One example of such an experiment is presented in Fig.~\ref{Fig3} for cross-linear polarization
of the resonant CW excitation and the two-photon detection of X$_2$-Mn. Without CW resonant excitation on X-Mn, an
identical intensisty is observed for the six lines
of X$_2$-Mn PL showing that the two-photon excitation
does not have any significant effect on the Mn spin
population. A strong change in the intensity distribution
is observed when the CW laser is scanned across each X-Mn
level suggesting a complex Mn spin dynamics.

Let us first focus on the two extreme lines in the X$_2$-Mn PL corresponding to the two degenerate spin states
$S_z=\pm5/2$ (linearly polarized detection). We observe a strong change in the intensity
distribution when the CW excitation laser is scanned across
the overall structure of X-Mn (Fig.~\ref{Fig3}(d)). For instance, each maximum in the
intensity of $S_z=\pm5/2$ occurs with a decrease in the
intensity of the other lines associated with $S_z=\pm3/2$
and $S_z=\pm1/2$ (doted blue lines in Fig.~\ref{Fig3}(d)). A CW excitation around each X-Mn level produces a modification of the Mn population distribution. For off-resonant excitation, the Mn
population is approximately equally distributed on the six spin states.

Fig.~\ref{Fig3}(b) presents two spectra of X$_2$-Mn obtained when
the CW excitation is scanned around the low energy line of
X-Mn $|J_z=\pm1,S_z=\mp5/2\rangle$. When the excitation is on
resonance with the low energy line, the population of
$S_z=\pm5/2$ is weaker than the others: This is the standard optical pumping regime. However, an increase of a few tens of $\mu$eV of the excitation energy
completely changes the Mn spin population distribution and most of the population can be transferred to states $S_z=\pm5/2$. PL spectra of X$_2$-Mn obtained in this excitation regime ($(ii)$ in Fig.~\ref{Fig3}(a)) with
and without transverse magnetic field ({\it i.e.} in the
QD plane) are presented in Fig.~\ref{Fig3}(c). A
transverse field B$_x$=0.22T almost completely restores an
equilibrium distribution in the PL intensities of the six
lines. This can be explained by the Mn spin precession induced by the transverse field, which completely erases any Mn spin memory. This magnetic field dependence confirms that the observed variations in the two-photon PL
of X$_2$-Mn are linked to the Mn spin population affected by the CW resonant laser. We will now analyse with this detection technique the Mn spin dynamics under resonant optical excitation.

\section{Mn spin population redistribution under resonant optical excitation}

\begin{figure}[hbt]
\includegraphics[width=3.5in]{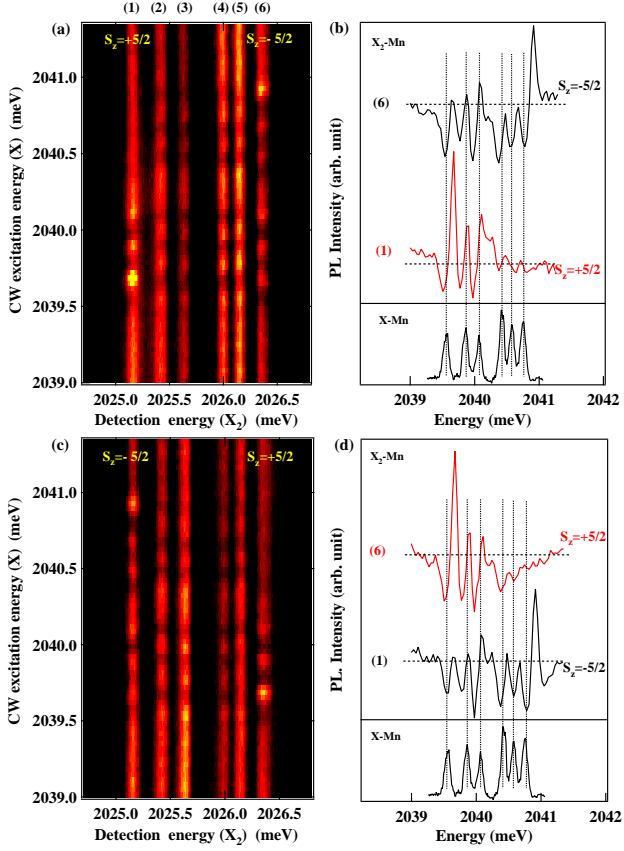}
\caption{(Color online) Map of the intensity of X$_2$-Mn in QD2 versus the
energy of the CW resonant excitation on X-Mn for co (a) and
cross (c) circular CW excitation and detection of X$_2$-Mn. (b) and (d)
present the corresponding intensity curves of X$_2$-Mn for
the spin states S$_z$=+5/2 (1) and S$_z$=-5/2 (6).} \label{Fig4}
\end{figure}

To address individual Mn spin states, resonant optical experiments
are performed under circularly polarized excitation and
detection. With circularly polarized photons, an excitation
on a given X-Mn level only affects one spin state of the Mn.
For co-circular excitation on X-Mn and detection on
X$_2$-Mn, the low energy lines of X-Mn and X$_2$-Mn
correspond to the same spin state of the Mn (see the level
scheme presented in Fig.~\ref{Fig1}). As presented in Fig.~\ref{Fig4}, when a CW excitation laser is
scanned around the low energy line of X-Mn, it mainly affects the low energy line of
X$_2$-Mn. Similarly, as the high energy line of X-Mn is
excited, the intensity of the high energy line of X$_2$-Mn
is significantly modified. Both configurations show that the resonant CW excitation mainly affects the spin state of the Mn which is resonantly excited.

To analyze the details of the influence of the CW resonant
laser on the Mn spin population, we focus on the two
outside lines of the PL of X$_2$-Mn. They correspond to the spin states $S_z=+5/2$ or $S_z=-5/2$. The intensity
of these lines is presented in Fig.~\ref{Fig4}(b) versus
the energy of the resonant CW laser. As expected
for an optical spin pumping mechanism, one observes a decrease
of the  S$_z$=+5/2 spin population when the laser is tuned on
resonance with the X-Mn level $|J_z=-1,S_z=+5/2\rangle$. However, a
strong increase of the S$_z$=+5/2 population is observed
when the CW control laser is slightly detuned on the
high energy side of the X-Mn optical transition. A similar
behavior is observed when the high energy line of X-Mn is
excited and the high energy line of X$_2$-Mn is probed
(i.e. exciting and detecting the Mn spin state S$_z$=-5/2).
This confirms that, as expected for a Mn spin dependent phenomena, reversing the polarization of detection from co-circular to cross-circular (Fig.\ref{Fig4}(c) and
(d)) reverses the role played by the high and low energy
lines of X$_2$-Mn: an identical strong increase of
population is observed on S$_z$=+5/2 or S$_z$=-5/2 for a
slightly detuned excitation on the corresponding X-Mn
states.

\begin{figure}[hbt]
\includegraphics[width=3.5in]{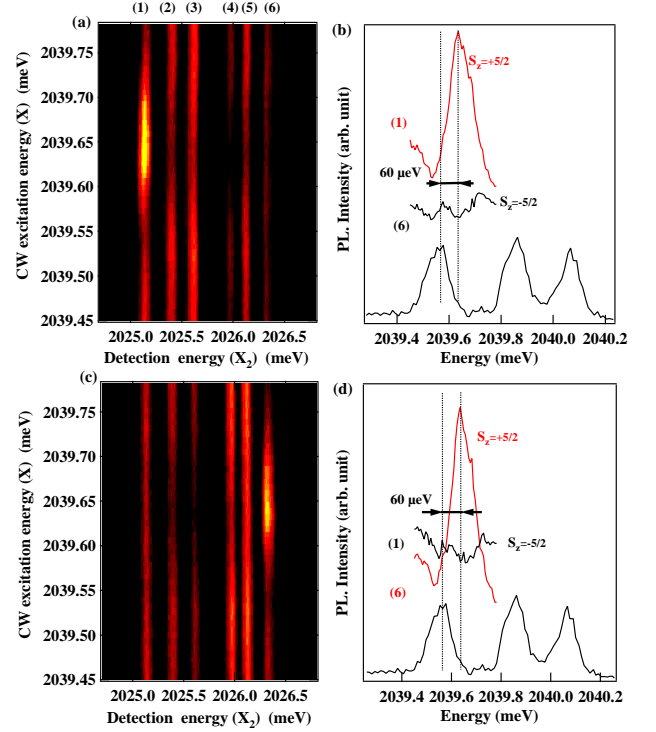}
\caption{(Color online) High resolution intensity map of X$_2$-Mn versus
the energy of a single mode resonant excitation scanned
around $|J_z=-1,S_z=+5/2\rangle$ for co (a) and cross (c) circular
excitation on X-Mn and detection on X$_2$-Mn (QD2). (b) and (d) present the
corresponding intensity curves of X$_2$-Mn for the spin
states S$_z$=-5/2 and S$_z$=+5/2.} \label{Fig5}
\end{figure}

A detail of the evolution of the PL intensity distribution
of X$_2$-Mn obtained by scanning a circularly polarized
CW laser around the low energy line of X-Mn is
presented in Fig.~\ref{Fig5}. These scans allow to
quantify the energy shift needed to optimize the Mn spin preparation. A maximum of the S$_z$=+5/2 spin
population is obtained for a laser detuning of about 60
$\mu eV$ when the excitation is scanned on the low or
high energy line of X$_2$-Mn in co or cross-circular
polarization respectively.

These experimental results show that the resonant CW excitation decreases the
population of the spin state which is resonantly excited as
expected from previous resonant optical pumping experiments
\cite{LeGall2010}. More surprisingly, a slight detuning of the resonant CW
laser increases significantly the Mn spin population in the state which
is optically addressed. We will see in the following that this
\emph{spin population trapping} is a specific signature of the coherent dynamics of the Mn
spin coupled with its nuclear spin and the resonant laser
field.

\section{Dynamics of a Mn atom under a resonant laser field}

To model the influence of a resonant laser field on the spin dynamics of a Mn atom embedded in a QD, we start from the analysis of the spin structure of a Mn atom in a strained zinc-blend semiconductor matrix. The Hamiltonian of
the coupled electronic and nuclear spins of a Mn atom in a strained layer grown along [001] axis is known from
magnetic resonance measurements \cite{Qazzaz1995} and reads:

\begin{eqnarray}
\label{MnStrain} {\cal H}_{Mn}= {\cal
A}\overrightarrow{I}.\overrightarrow{S}
\nonumber\\
+\frac{1}{6}a[S_x^4+S_y^4+S_z^4-\frac{1}{5}S(S+1)(3S^2+3S-1)]
\nonumber\\
+{\cal D}_0[S_z^2-\frac{1}{3}S(S+1)]+E[S_x^2-S_y^2]
\nonumber\\
+g_{Mn}\mu_B\overrightarrow{B}.\overrightarrow{S}
\end{eqnarray}

\noindent where ${\cal A}$ is the hyperfine coupling (${\cal A}\approx+0.7\mu eV$)
\cite{Causa1980}, which results from the magnetic dipolar interaction between
the Mn 5$d$ electrons forming the total spin $\vec{S}$ and
the spin of the Mn nucleus $\vec{I}$ (I=5/2). The second term of the Hamiltonian
comes from the cubic symmetry of the crystal field and mixes different
$S_z$ of the Mn spin. We have $a=0.32\mu eV$ according to reference
25.

The presence of bi-axial strains in the QD plane leads to the magnetic anisotropy term with ${\cal D}_{0}\approx12\mu eV$ for a fully strained CdTe layer matched on a ZnTe substrate. Because of partial relaxation of the strain during the growth process, weaker values of ${\cal D}_{0}$ are usually observed in self-assembled QDs \cite{LeGall2009}. An anisotropy of the strain in the $xy$ plane (QD plane) can mix different S$_z$ components through the anisotropic crystal field. This coupling is described in the Hamiltonian (1) by its characteristic energy $E$ which depends on the local strain distribution at the Mn atom location.

In highly strained QDs at zero magnetic field, the Mn
electronic spin is quantized along the growth axis $z$ and the
different electronic spin doublets (S$_z$=$\pm$1/2,
S$_z$=$\pm$3/2 and S$_z$=$\pm$5/2) are separated by an
energy proportional to ${\cal D}_0$ (Fig. 6). The S$_z$=$\pm$5/2
and S$_z$=$\pm$3/2 doublets are split into six lines by the
hyperfine coupling with the nuclear spin I=5/2. For the
doublet S$_z$=$\pm$1/2, the isotropic coupling with the
nuclear spin I=5/2 results in two levels with total spin
M=2 and M=3. The regular spacing between the electronic
spin levels can be restored by a magnetic field B$\geq$0.5T (last term of Hamiltonian (1)).

\begin{figure}[hbt]
\includegraphics[width=3in]{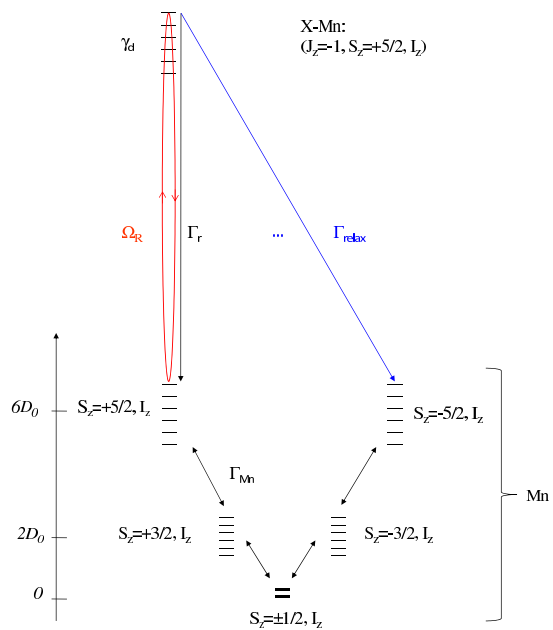}
\caption{(Color online) Scheme of the energy levels and transitions rates
involved in the resonant excitation model (see text). $\hbar\Omega_R$ is
the energy coupling with the laser, $\gamma_d=1/\tau_d$ is a pure dephasing
rate of the exciton, $\Gamma_{Mn}=1/\tau_{Mn}$ is the spin relaxation
rate of the Mn, $\Gamma_{r}=1/\tau_{r}$ is the optical recombination
rate of the exciton. A relaxation rate of the exciton-Mn
complex $\Gamma_{relax}=1/\tau_{relax}$ is used for an effective
description of the optical pumping effect.}
\label{Fig6}
\end{figure}

Different S$_z$ of the Mn are coupled by the non-diagonal
terms of $H_{Mn}$ (equation (1)). For instance,
the hyperfine terms ${\cal A}$ couples two consecutive Mn
spin states through an electron-nuclei flip-flop. We could
then expect that a non-equilibrium population of the
electronic spin prepared optically would be transferred to the nuclear spin.
This would lead to an optical pumping of the nuclear spin of the Mn. However, in the presence of a
large magnetic anisotropy, these electron-nuclei flip-flops
are blocked.

An anisotropic strain distribution in the QD
plane can also efficiently couple Mn spin states S$_z$ separated by
two units through the crystal field term $E(S_x^2-S_y^2)$.
As we will see, all these coupling terms affect the population redistribution on the
six Mn spin states under resonant optical pumping.

The coupling with a resonant laser field can be used
to tune the energy of one selected Mn spin state across the
full fine structure of the Mn atom \cite{LeGall2011}. This allows in particular to restore
the degeneracy of two consecutive Mn spins states giving an
optical way to control the flip-flops of the electronic and
nuclear spins. The energy tuning of the optically dressed
states is then expected to influence the dynamics of the
coupled electronic and nuclear spins. To estimate this
effect, we calculated the coherent evolution of coupled
electronic and nuclear spins optically coupled to an
X-Mn state.

As illustrated in Fig.~\ref{Fig6}, we first consider that a single
exciton state ($|J_z=-1\rangle$) is laser coupled to one
state of the Mn ($|S_z=+5/2,I_z\rangle$) with a Rabi energy
$\hbar\Omega_R$. This approximation is justified in strongly
confined QDs with a large X-Mn splitting resonantly excited
by a narrow band laser and in the limit of small laser detuning. The exciton has a
pure dephasing rate $\gamma_d$ and the relaxation of the Mn
spin in the ground state  (empty QD) is described by a relaxation
rate $\Gamma_{Mn}$ coupling one by one the different electronic spin
states $S_z$. The nuclear spin $I_z$ is considered to be frozen in the
timescale of all the spin preparation mechanism discussed here.

The X-Mn complex can relax its energy along a Mn spin conserving channel at rate $\Gamma_r$ (optical recombination of X) or along channels including a relaxation of the Mn spin at rate $\Gamma_{relax}=1/\tau_{relax}$: $\Gamma_{relax}$ allows a transfer of population from the state $|J_z=-1,S_z=+5/2,I_z\rangle$ to any other spin state of
the Mn $S_z$ with $I_z$ unchanged. This is a
simplified effective way to describe the complex X-Mn spin
dynamics at the origin of the optical pumping mechanism \cite{Cao2011,Cywinski2010}. At
magnetic fields lower than a few hundreds mT, we also consider that the Zeeman energy of the X-Mn can be neglected since it is much smaller than the X-Mn exchange interaction: we only take into account the effect of the magnetic  field on the empty QD (Last term of Hamiltonian (1) for a Mn alone).

Using the simplified level scheme presented in Fig.\ref{Fig6}, we can calculate the time evolution of the 42x42 density matrix $\varrho$ describing the population and the coherence of the 36 states of the Mn alone (empty QD described by ${\cal H}_{Mn}$) and the 6 X-Mn states $|J_z=-1,S_z=+5/2,I_z\rangle$. The master equation which governs the evolution of $\varrho$ can be written in a general form (Lindblad form) as:

\begin{equation}
\label{Lindblad} {\frac{\partial\varrho}{\partial
t}=-i/\hbar[{\cal H},\varrho]+L\varrho}
\end{equation}

\noindent ${\cal H}$ is the Hamiltonian of the complete system (Mn and X-Mn) and $L\varrho$ describes the coupling or decay channels resulting from an
interaction with the environment \cite{Exter2009}. One can split $L\varrho$ in three parts:

1- The population transfer from level $j$ to level $i$ in an irreversible process associated with a
coupling to a reservoir is described by $L_{inc,j\rightarrow
i}\varrho$:

\begin{eqnarray}
\label{inc} { L_{inc,j\rightarrow
i}\varrho=\frac{\Gamma_{j\rightarrow
i}}{2}(2|i\rangle\langle j|\varrho|j\rangle\langle i|
-\varrho|j\rangle\langle j|-|j\rangle\langle j|\varrho)}
\end{eqnarray}

\noindent where $\Gamma_{j\rightarrow i}$ is the incoherent
relaxation rate from level $j$ to level $i$. This operator describes the radiative decay of the exciton
(irreversible coupling to the photon modes) or the
relaxation of the Mn spin (irreversible coupling to the
phonon modes). Such term could also be used to describe the
optical generation of an exciton in the low excitation
regime where the energy shift induced by the strong coupling with the laser field is neglected.

2- A pure dephasing (\emph{i.e.} not related to an exchange
of energy with a reservoir) is also introduced for the exciton and described by $L_{deph,jj}\varrho$:

\begin{eqnarray}
\label{deph} {
L_{deph,jj}\varrho=\frac{\gamma_{jj}}{2}(2|j\rangle\langle
j|\varrho|j\rangle\langle j| -\varrho|j\rangle\langle
j|-|j\rangle\langle j|\varrho)}
\end{eqnarray}

\noindent where $\gamma_{jj}$ is a pure dephasing rate.

3- For a general description, valid from the low to the
high optical excitation intensity regime, we consider that the
laser field induces a coherent coupling between the ground
and exciton states. The coherent coupling between two
levels induced by the laser field leads to Rabi oscillations
between the populations $\varrho_{ii}$ and $\varrho_{jj}$
and coherence of these levels
$\varrho_{ij}=\varrho_{ji}^*$. In the Lindblad equation
(\ref{Lindblad}), this reversible coupling can be described
by $L_{coh,i\leftrightarrow j}$:

\begin{eqnarray}
\label{coh} { L_{coh,i\leftrightarrow
j}\varrho=i\frac{\Omega_{ij}}{2}(|j\rangle\langle
i|\varrho+|i\rangle\langle
j|\varrho-\varrho|j\rangle\langle
i|-\varrho|i\rangle\langle j|)}
\end{eqnarray}

\noindent where $\hbar\Omega_{ij}={\cal P}_{ij}{\cal E}$ is the Rabi energy
splitting with ${\cal P}_{ij}$ the dipolar moment of the QD transition and ${\cal E}$ the amplitude of the electric field
of the resonant CW laser. This term, which corresponds to the
dipole-field coupling $-\overrightarrow{{\cal P}_{ij}}.\overrightarrow{{\cal E}}=-\hbar\Omega_{ij}(|j\rangle\langle
i|+|i\rangle\langle j|)/2$, could also be included in the Hamiltonian
evolution (first term of equation (\ref{Lindblad}))\cite{Exter2009,Roy2011}.

\begin{figure}[hbt]
\includegraphics[width=3.5in]{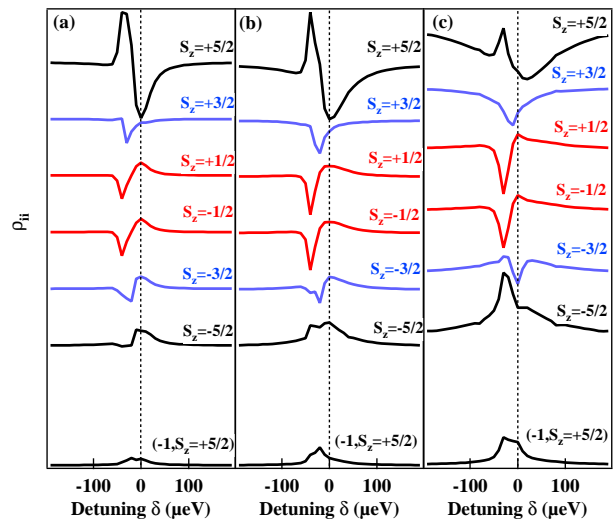}
\caption{(Color online) Calculated population of the six electronic spin states of a
Mn versus the detuning $\delta$ of the CW control laser around the
X-Mn state $|J_z=-1,S_z=+5/2,I_z\rangle$ for different Rabi
energies: (a) $\hbar\Omega_R=12.5\mu eV$, (b) $\hbar\Omega_R=25\mu eV$
and (c) $\hbar\Omega_R=50\mu eV$. The relaxation times are $\tau_{Mn}=250ns$, $\tau_r=0.25 ns$, $\tau_{relax}=60 ns$, $\tau_d=100 ps$ and the Mn fine structure parameters ${\cal D}_{0}=7\mu eV$ and $E=0.35\mu eV$.} \label{Fig7}
\end{figure}

The calculated evolution of the population of the different
spin states of the Mn with the detuning of a circularly
polarized laser around $|J_z=-1,S_z=+5/2,I_z\rangle$ is presented in
Fig.~\ref{Fig7} for different Rabi energies. The detuning $\delta$ is defined as $\delta=\hbar\omega_0-\hbar\omega_l$ with $\hbar\omega_0$ the energy of the optical excitonic transition and $\hbar\omega_l$ the energy of the CW laser \cite{Exter2009}. The states
$|S_z=+5/2,I_z\rangle$ of the Mn are coupled by the CW resonant
laser to the X-Mn states $|J_z=-1,S_z=+5/2,I_z\rangle$ and
we neglect the possible excitation of more than one exciton
state by the resonant laser (i.e. the splitting between the
X-Mn lines is larger than the Rabi energy or the detuning
of the laser). At low Rabi energies and zero
detuning, a strong decrease of the population of S$_z$=+5/2 and an increase of the population of the other spin states is observed: this corresponds to the expected optical pumping of the state S$_z$=+5/2. As the laser is slightly detuned on the high energy side of the transition, $\delta<0$, a strong increase of the
S$_z$=+5/2 population and simultaneous decrease of the  S$_z$=$\pm$1/2 population is observed. This detuning dependence, that we call \emph{spin population trapping}, is very similar to the experimental data.

\begin{figure}[hbt]
\includegraphics[width=3.5in]{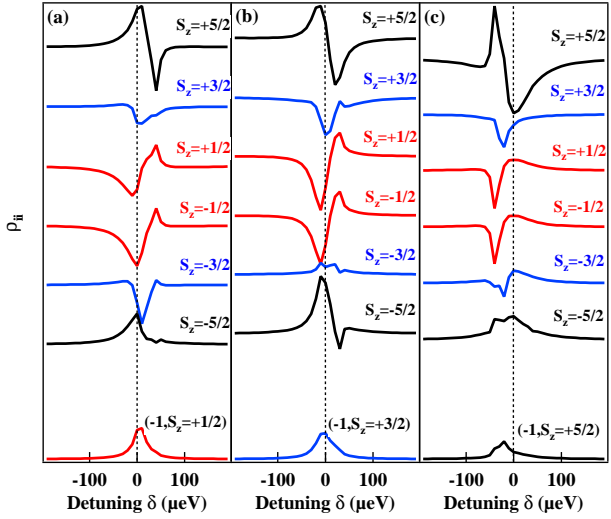}
\caption{(Color online) Calculated population of the six spin states of the
Mn versus the detuning $\delta$ of the CW laser around the
X-Mn states $|J_z=-1,S_z=+1/2,I_z\rangle$ (a),
$|J_z=-1,S_z=+3/2,I_z\rangle$ (b) and $|J_z=-1,S_z=+5/2,I_z\rangle$ (c)
at a fixed Rabi energy $\hbar\Omega_R=25\mu eV$. The relaxation times are $\tau_{Mn}=250ns$, $\tau_r=0.25 ns$, $\tau_{relax}=60 ns$, $\tau_d=100 ps$ and the Mn fine structure parameters ${\cal D}_{0}=7\mu eV$ and $E=0.35\mu eV$.}
\label{Fig8}
\end{figure}

The calculation shows that the states S$_z$=+5/2
(excited by the CW resonant laser) and S$_z$=$\pm$1/2 are
the most affected by the laser detuning (see for instance Fig.~\ref{Fig7}(b)). Let us give a
qualitative description of the observed complexe spin dynamics.
As the CW laser is detuned on the high energy side of the transition, the optically
dressed states associated with S$_z$=+5/2 can be pushed on
resonance with S$_z$=$\pm$1/2. At resonance, mixed states of S$_z$=+1/2 and S$_z$=+5/2 are created through the anisotropic crystal field (E term of equation (1)). This coherent coupling produces
an enhancement of the population transfer between
S$_z$=+1/2 and the optically dressed states associated with S$_z$=+5/2.

The optical recombination of the optically dressed state, which
is mainly an excitonic state, induces an irreversible transfer of
population from S$_z$=+1/2 to S$_z$=+5/2. In addition,
S$_z$=+1/2 and S$_z$=-1/2 are coherently coupled by the
hyperfine interaction which, through electron-nuclei
flip-flops, produces an oscillation of population between
these two levels. This oscillation is interrupted by the irreversible transfer of S$_z$=+1/2 to the optically dressed states. Consequently, both the S$_z$=+1/2 and the
S$_z$=-1/2 populations are transferred to S$_z$=+5/2. This mechanism can explain the strong increase of the S$_z$=+5/2 (or S$_z$=-5/2) population for a circularly polarized CW laser excitation slightly detuned around the high or the low energy line of X-Mn.

\begin{figure}[hbt]
\includegraphics[width=3.5in]{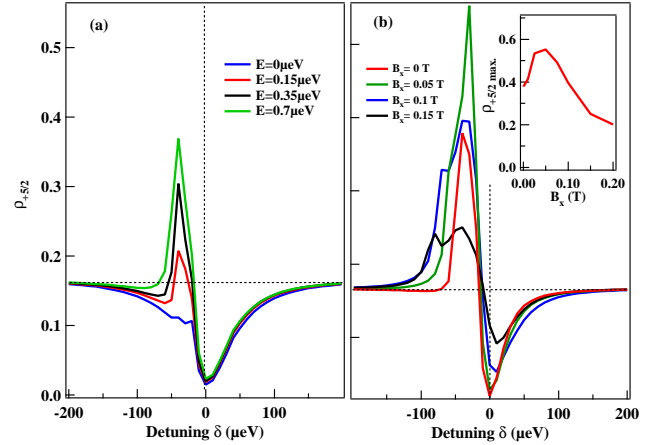}
\caption{(Color online) (a) Calculated population of S$_z$=+5/2 versus the
CW laser detuning, $\delta$, around $|J_z=-1,S_z=+5/2,I_z\rangle$: (a) for different values of the in-plane anisotropy of the strain $E$ at a fixed Rabi energy $\hbar\Omega_R=25\mu eV$ and (b) for differrent transverse
magnetic field, $B_{x}$, with E=$0.35\mu eV$ and $\hbar\Omega_R=12.5\mu eV$. The inset shows
the transverse magnetic field dependence of the peak
population of S$_z$=+5/2. The relaxation times are $\tau_{Mn}=250ns$, $\tau_r=0.25 ns$, $\tau_{relax}=60 ns$, $\tau_d=100 ps$ and the magnetic anisotropy ${\cal D}_{0}=7\mu eV$.} \label{Fig9}
\end{figure}

The spectral width of the pumping signal (i.e. decrease of population obtained on the
resonance with the transition) increases with the Rabi
energy. Consequently, the optical pumping significantly affects the population trapping mechanism and a more complex dynamics is expected at hight
excitation intensity (calculation presented in Fig.~\ref{Fig7}(c) for $\hbar\Omega_R=50\mu eV$, larger than the excitation intensity used in the experiments presented here). In this high excitation regime, the population of S$_z$=$\pm$1/2 is transferred to S$_z$=+5/2 by the population trapping mechanism and simultaneously, the optical pumping empties S$_z$=+5/2. This leads to a transfer from S$_z$=+5/2 and S$_z$=$\pm$1/2 to S$_z$=-3/2 and S$_z$=-5/2.

In Fig.~\ref{Fig8}, we report the population of the six spin states of the Mn at fixed Rabi
energy $\hbar\Omega_r=25\mu eV$ for CW laser detunings around three X-Mn levels: $|J_z=-1,S_z=+1/2\rangle$,
$|J_z=-1,S_z=+3/2\rangle$ and $|J_z=-1,S_z=+5/2\rangle$. For each X-Mn level, a
strong deviation from the standard optical pumping regime
is observed, in agreement with the behaviour observed in the experiments presented in Fig.~\ref{Fig3}. A general trend of these simulations
is that the population of S$_z$=+1/2 and S$_z$=-1/2 always
have the same laser detuning dependence. These two states are mixed by the hyperfine coupling with the nuclear spin in an empty dot (Mn alone) and their populations tend to equilibrate as soon as the exciton recombines, whatever the excitation condition are. An excitation on S$_z$=+1/2 also
strongly affects the population of S$_z$=+5/2. This is
another signature of the coherent coupling between these states induced by the
strained anisotropy term $E$. When the dressed state of
S$_z$=+1/2 is on resonance with S$_z$=+5/2 ($\delta>0$), the population
is transferred to S$_z$=+1/2 strongly coupled with
S$_z$=-1/2, this is similar to the population trapping mechanism
discussed in the case of an excitation on S$_z$=+5/2.

The situation is more complex for an excitation on S$_z$=+3/2 as the standard optical pumping and the population transfer to the optically dressed state can occur simultaneously. The dressed states of S$_z$=+3/2 can be tuned on resonance with the state S$_z$=+5/2 (positive detuning) or with the state S$_z$=+1/2 (negative detuning) for small laser detuning where an efficient optical pumping can take place. For a small negative detuning, an efficient transfer towards the dressed states of S$_z$=+3/2 from S$_z$=+1/2 and S$_z$=-1/2 occurs via flip-flops with the nuclear spin. Simultaneously, the optical pumping empties S$_z$=+3/2. This leads to a transfer of population from S$_z$=+3/2, S$_z$=+1/2 and S$_z$=-1/2 to S$_z$=$\pm$5/2. This process involves electron nuclei flip-flops and it is expected to take place together with an optical pumping of the Mn nuclear spin. For a positive detuning, a transfer of population occurs
from S$_z$=+5/2 to S$_z$=+3/2 via flip-flops with the nuclear spin. The optical pumping tends to empty S$_z$=+3/2 and the population is mainly transferred to S$_z$=$\pm$1/2.

The in-plane strain anisotropy $E$ plays a major role in the population trapping mechanism presented here. This is confirmed by the calculation displayed in Fig.~\ref{Fig9}(a). In the absence of in-plane strain anisotropy ($E$=0) the population trapping disappears and the detuning dependence of the Mn spin
population is dominated by the standard optical spin pumping mechanism. A
value of $E$ corresponding to a few percent of ${\cal D}_0$ (5$\%$
in the calculation presented in Fig.~\ref{Fig7}) can
explain the characteristic redistribution of population
observed in the resonant excitation experiments.

These calculated detuning dependence explain the non-trivial variation of intensities observed when scanning a CW laser across the overall X-Mn spectrum. Excitation on any Mn spin state mainly affects the population of $S_z\pm1/2$ and $S_z\pm5/2$ and the detuning dependence of the Mn spin population always deviates from the standard resonant optical pumping regime.

\begin{figure}[hbt]
\includegraphics[width=3.2in]{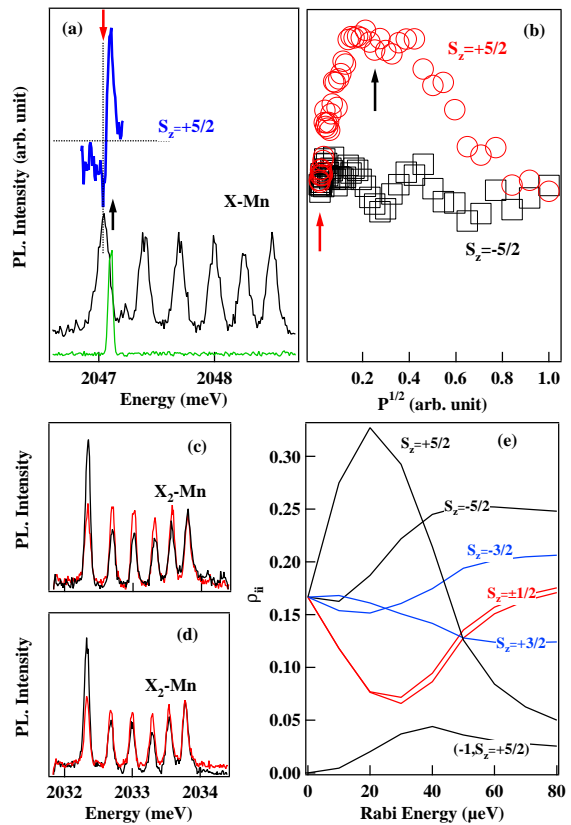}
\caption{(Color online) (a) PL of X-Mn in QD3 (black). The green curve is
the spectra of the single mode laser used for the power
dependence presented in (b). The bleu curve presents the
intensity of X$_2$-Mn for the spin state S$_z$=+5/2 versus
the energy of the $\sigma$- single mode laser scanned
around the low energy line of X-Mn. (b) PL intensity of the high (S$_z$=-5/2, black squares) and
low (S$_z$=+5/2, red circles) energy lines of X$_2$-Mn versus the square root of the intensity P of the resonant CW laser (quantity proportional to the laser electric field). (c)
PL spectra of X$_2$-Mn for two slightly different
excitation energies: excitation energy pointed by the red
arrow in (a) (red curve) and by the black arrow in (a)
(black curve). (d) PL  spectra of X$_2$-Mn for two
different excitation power: excitation power pointed by the
red arrow in (b) (red curve) and by the black arrow in (b)
(black curve). (e) Calculated power dependence of the
different spin states populations for an excitation detuned
by $\delta=-40\mu eV$ from $|J_z=-1,S_z=+5/2,I_z\rangle$. The parameters used in the calculation are $\tau_{Mn}=250ns$, $\tau_r=0.25 ns$, $\tau_{relax}=60 ns$, $\tau_d=100 ps$, ${\cal D}_{0}=7\mu eV$ and E = 0.35$\mu
eV$.}
\label{Fig10}
\end{figure}

Since the population trapping relies on the coherent coupling between the Mn spin states, one may wonder how to tune it in order to optimize the spin preparation. One way to control this coupling is by applying a small external magnetic field. As illustrated in Fig.~\ref{Fig9}(b) a weak in-plane magnetic field induces coherent coupling among the Mn spin states and enhances the population trapping. However, as already observed in optical pumping experiments \cite{LeGall2009,LeGall2010}, an equilibrium of the different S$_z$ states population is restored when the in-plane magnetic field overcomes the magnetic anisotropy independently of the optical excitation conditions, in agreement with the experimental data presented in Fig.~\ref{Fig3}(c). In the present case, the spin population trapping - and consequently the fidelity of the Mn initialization - is maximum for a transverse magnetic field around 0.05 T.

In addition to the CW laser detuning, its  power is another way to control the Mn spin preparation. In the calculated excitation power dependence presented in Fig.~\ref{Fig10}(e), for a detuning between the pumping laser and the excited transition $|J_z=-1,S_z=+5/2\rangle$ of $\delta$=-40$\mu$eV, even at low excitation intensity the optically dressed state is on resonance with S$_z$=$\pm$1/2. The optical trapping regime already starts at very low excitation intensity. As the excitation intensity increases, the dressed state is shifted away from the S$_z$=$\pm$1/2 spin states and the population trapping
disappears. A standard optical pumping of the state
S$_z$=+5/2 is restored. This non-monotonic evolution of the
population of the resonantly excited level is characteristic
of the regime of optical spin population trapping.

A non-monotonic power dependence is indeed observed in the
experiments presented in Fig.~\ref{Fig10}(b). For a single
mode excitation laser detuned on the high energy side of the state $|J_z=-1,S_z=+5/2\rangle$ (Fig.~\ref{Fig10}(a)), an increase of the population is observed
starting from the low excitation intensity regime. This is the spin
population trapping regime. Then the population saturates
and decreases at high excitation intensity when the optically dressed
state is shifted away from the states S$_z$=$\pm$1/2 and
the transfer of population is suppressed. Examples of low and high
excitation power spectrum of X$_2$-Mn are presented in
Fig.~\ref{Fig10}(d). This experiment confirms that the optical spin population trapping regime
can be reached either by changing the laser detuning or by
changing the laser excitation intensity.

\section{Conclusion}

In conclusion, we have shown that the coupling with a resonant
laser field strongly modifies the spin dynamics of a
Mn atom inserted in a strained self-assembled QD. In addition to the standard optical pumping, the Mn spin can be trapped in the state which is optically excited. This mechanism of \emph{spin population trapping} is controlled by the presence of a coherent coupling between the different Mn spin states S$_z$ induced by an in-plane strain anisotropy. Such spin dynamics is not specific to a Mn atom and could be observed in other solid state and atomic spin systems provided that a coherent coupling between the spin sub-levels is present. As demonstrated for a Mn atom, the coherent coupling could be induced by a transverse magnetic field.

The population trapping of the Mn electronic spin can also involve flip-flops with the Mn nuclear spin and an optical pumping of the nuclear spin is expected. This optical excitation configuration will be used in future experiments to optically access the nuclear spin of the Mn in its solid state environment. This spin initialization technique could also be extended to QDs containing 2 Mn atoms \cite{Besombes2012} to optically induce a correlation between the localized spins.

\begin{acknowledgements}
This work is supported by the French ANR contract QuAMOS
and the EU ITN contract Spin-Optronics.
\end{acknowledgements}


\begin{thebibliography}{}


\bibitem{Brossel1952} J. Brossel and F. Bitter, Phys. Rev. {\bf 86}, 17308 (1952).
\bibitem{Atature2006} M. Atatüre, J. Dreiser, A. Badolato, A. Högele, K. Karrai and A. Imamoglu, Science {\bf 312}, 551 (2006).
\bibitem{Gerardot2008} B. D. Gerardot, D. Brunner, P. A. Dalgarno, P. Ohberg, S. Seidl, M. Kroner, K. Karrai, N. G. Stoltz, P. M. Petroff and R. Warburton, Nature {\bf 451}, 441 (2008).
\bibitem{LeGall2010} C. Le Gall, R. S. Kolodka, C. L. Cao, H. Boukari, H. Mariette, J. Fernandez-Rossier and L. Besombes, Phys. Rev. B {\bf 81}, 245315 (2010).
\bibitem{Mollow1972} B. R. Mollow, Phys. Rev. A 5, 2217 (1972).
\bibitem{Jundt2008} G. Jundt, L. Robledo, A. Hogele, S. Falt and A. Imamoglu, Phys. Rev. Lett. {\bf 100}, 177401 (2008).
\bibitem{Xu2007} X. Xu, B. Sun, P.R. Berman, D.G. Steel, A.S. Bracker, D. Gammon and L.J. Sham, Science {\bf 317}, 929 (2007).
\bibitem{Xu2008} X. Xu, Bo Sun,  E. D. Kim, K. Smirl, P.R. Berman, D.G. Steel, A.S. Bracker, D. Gammon and L.J. Sham, Phys. Rev. Lett. {\bf 101}, 227401  (2008).
\bibitem{Kroner2008} M. Kroner, C. Lux, S. Seidl, A. W. Holleitner, K. Karrai, A. Badolato, P. M. Petroff and R.J. Warburton, Appl. Phys. Lett. {\bf 92}, 031108 (2008).
\bibitem{Muller2008} A. Muller, W. Fang, J. Lawall and G.S. Solomon, Phys. Rev. Lett. {\bf 101}, 027401 (2008).
\bibitem{Besombes2004} L. Besombes, Y. Leger, L. Maingault, D. Ferrand, H. Mariette and J. Cibert, Phys. Rev. Lett. {\bf 93}, 207403 (2004).
\bibitem{Kudelski2007} A. Kudelski, A. Lemaitre, A. Miard, P. Voisin, T.C.M. Graham, R.J. Warburton and O. Krebs, Phys. Rev. Lett. {\bf 99}, 247209 (2007).
\bibitem{LeGall2011} C. Le Gall, A. Brunetti, H. Boukari and L. Besombes, Phys. Rev. Lett. {\bf 107}, 057401 (2011).
\bibitem{Wojnar2011} P. Wojnar, C. Bougerol, E. Bellet-Amalric, L. Besombes, H. Mariette and H. Boukari, J. Crystal Growth {\bf 335}, 28 (2011).
\bibitem{Trojnar2012} A. H. Trojnar, M. Korkusinski, M. Potemski and P. Hawrylak, Phys. Rev. B {\bf 85}, 165415 (2012).
\bibitem{Fernandez2006} J. Fernandez-Rossier, Phys. Rev. B {\bf 73}, 045301 (2006).
\bibitem{LeGall2009} C. Le Gall, L. Besombes, H. Boukari, R. Kolodka, J. Cibert and H. Mariette, Phys. Rev. Lett. {\bf 102}, 127402 (2009).
\bibitem{Goryca2009} M. Goryca, T. Kazimierczuk, M. Nawrocki, A. Golnik, J. A. Gaj, P. Kossacki, P. Wojnar and G. Karczewski, Phys. Rev. Lett. {\bf 103}, 087401 (2009).
\bibitem{Reiter2009} D.E. Reiter, T. Kuhn and V.M. Axt, Phys. Rev. Lett. {\bf 102}, 177403 (2009).
\bibitem{Reiter2012} D.E. Reiter, T. Kuhn and V.M. Axt, Phys. Rev. B {\bf 85}, 045308 (2012).
\bibitem{Besombes2008} L. Besombes, Y. Leger, J. Bernos, H. Boukari, H. Mariette, J.P. Poizat, T. Clement, J. Fernandez-Rossier and R. Aguado, Phys. Rev. B {\bf 78} 125324 (2008).
\bibitem{Besombes2005} L. Besombes, Y. Leger, L. Maingault, D. Ferrand, H. Mariette and J. Cibert, Phys. Rev. B {\bf 71}, 161307 (2005).
\bibitem{Flissikowski2004} T. Flissikowski, A. Betke, I.A. Akimov and F. Henneberger, Phys. Rev. Lett. {\bf 92}, 227401 (2004).
\bibitem{Qazzaz1995} M. Qazzaz, G. Yang, S.H. Xin, L. Montes, H. Luo and J.K. Furdyna, Solid State Communications {\bf 96}, 405 (1995).
\bibitem{Causa1980} M.T. Causa, M. Tovar, S.B. Oseroff, R. Calvo and W. Giriat, Phys. Lett. {\bf A77}, 473 (1980).
\bibitem{Cywinski2010} L. Cywinski, Phys. Rev. B {\bf 82}, 075321 (2010).
\bibitem{Cao2011} C. L. Cao, L. Besombes and J. Fernandez-Rossier, Phys. Rev. B {\bf84}, 205305 (2011).
\bibitem{Exter2009} M.P. van Exter, J. Gudat, G. Nienhuis and D. Bouwmeester, Phys. Rev. A {\bf 80}, 023812 (2009).
\bibitem{Roy2011} C. Roy and S. Hughes, Phys. Rev. X {\bf 1}, 021009 (2011).
\bibitem{Besombes2012} L. Besombes, C.L. Cao, S. Jamet, H. Boukari and J. Fernandez-Rossier, Phys. Rev. B 86, 165306 (2012).

\end{thebibliography}
\end{document}